# Probabilistic Frequent Pattern Growth for Itemset Mining in Uncertain Databases
## (Technical Report)

Thomas Bernecker, Hans-Peter Kriegel, Matthias Renz, Florian Verhein*
and Andreas Züfle

{*bernecker,kriegel,renz,verhein,zuefle*}@dbs.ifi.lmu.de

Institute for Informatics, Ludwig-Maximilians-Universität München, Germany

**Abstract.** Frequent itemset mining in uncertain transaction databases semantically and computationally differs from traditional techniques applied on standard (certain) transaction databases. Uncertain transaction databases consist of sets of existentially uncertain items. The uncertainty of items in transactions makes traditional techniques inapplicable. In this paper, we tackle the problem of finding probabilistic frequent itemsets based on possible world semantics. In this context, an itemset $X$ is called frequent if the *probability* that $X$ occurs in at least *minSup* transactions is above a given threshold $\tau$. We make the following contributions: We propose the first probabilistic FP-Growth algorithm (ProFP-Growth) and associated probabilistic FP-Tree (ProFP-Tree), which we use to mine all probabilistic frequent itemsets in uncertain transaction databases without candidate generation. In addition, we propose an efficient technique to compute the support probability distribution of an itemset in linear time using the concept of generating functions. An extensive experimental section evaluates the our proposed techniques and shows that our ProFP-Growth approach is significantly faster than the current state-of-the-art algorithm.

## 1 Introduction

Association rule analysis is one of the most important fields in data mining. It is commonly applied to market-basket databases for analysis of consumer purchasing behavior. Such databases consist of a set of transactions, each containing the items a customer purchased. The most important and computationally intensive step in the mining process is the extraction of *frequent itemsets* – sets of items that occur in at least *minSup* transactions. It is generally assumed that the items occurring in a transaction are known for certain. However, this is not always the case. For instance;

– In many applications the data is inherently noisy, such as data collected by sensors or in satellite images.

---

* Contact author. *verhein@dbs.ifi.lmu.de* or *http://www.florian.verhein.com/contact/*

- In privacy protection applications, artificial noise can be added deliberately [19]. Finding patterns despite this noise is a challenging problem.
- By aggregating transactions by customer, we can mine patterns across customers instead of transactions. This produces estimated purchase probabilities per item per customer rather than certain items per transaction.

In such applications, the information captured in transactions is *uncertain* since the existence of an item is associated with a likelihood measure or existential probability. Given an uncertain transaction database, it is not obvious how to identify whether an item or itemset is frequent because we generally cannot say for certain whether an itemset appears in a transaction. In a traditional (certain) transaction database on the other hand, we simply perform a database scan and count the transactions that include the itemset. This does not work in an uncertain transaction database.

An example of a small uncertain transaction database is given in Figure 1, where for each transaction $t_i$, each item $x$ is listed with its probability of existing in $t_i$. Items with an existential probability of zero can be omitted. We will use this dataset as a running example.

Prior to [6], expected support was used to deal with uncertain databases [7,8]. It was shown in [6] that the use of expected support in probabilistic databases had significant drawbacks which led to misleading results. The proposed alternative was based on computing the entire probability distribution of itemsets' support, and achieved this in the same runtime as the expected support approach by employing the Poisson binomial recurrence relation. [6] adopts an Apriori-like approach, which is based on an *anti-monotone* Apriori property [3] (if an itemset $X$ is not frequent, then any itemset $X \cup Y$ is not frequent) and candidate generation.

However, it is well known that Apriori-like algorithms suffer a number of disadvantages. First, all candidates generated must fit into main memory and the number of candidates can become prohibitively large. Secondly, checking whether a candidate is a subset of a transaction is non-trivial. Finally, the entire database needs to be scanned multiple times. In uncertain databases, the effective transaction width is typically larger than in a certain transaction database which in turn can increase the number of candidates generated and the resulting space and time costs.

In certain transaction databases, the FP-Growth Algorithm [11] has become the established alternative. By building an FP-Tree – effectively a compressed and highly indexed structure storing the information in the database – candidate generation and multiple database scans can be avoided. However, extending this idea to mining probabilistic frequent patterns in uncertain transaction databases is non-trivial. It should be noted that previous extensions of FP-Growth to uncertain databases used the expected support approach [1,14]. This is much easier since these approaches ignore the probability distribution of support.

In this paper, we propose a compact data structure called the probabilistic frequent pattern tree (*ProFP-tree*) which compresses probabilistic databases and allows the efficient extraction of the existence probabilities required to compute the support probability distribution and frequentness probability. Additionally, we propose the novel *ProFPGrowth* algorithm for mining all probabilistic frequent itemsets without candidate generation.

| TID | Transaction |
|-----|-------------|
| 1 | (A, 1.0), (B, 0.2), (C, 0.5) |
| 2 | (A, 0.1), (D, 1.0)) |
| 3 | (A, 1.0), (B, 1.0), (C, 1.0), (D, 0.4) |
| 4 | (A, 1.0), (B, 1.0), (D, 0.5) |
| 5 | (B, 0.1), (C, 1.0) |
| 6 | (C, 0.1), (D, 0.5) |
| 7 | (A, 1.0), (B, 1.0), (C, 1.0) |
| 8 | (A, 0.5), (B, 1.0) |

**Fig. 1.** Uncertain Transaction Database (running example)

### 1.1 Uncertain Data Model

The uncertain data model applied in this paper is based on the possible worlds semantic with existential *uncertain items*.

**Definition 1** *An* uncertain item *is an item $x \in I$ whose presence in a transaction $t \in T$ is defined by an* existential probability $P(x \in t) \in (0, 1)$. *A* certain item *is an item where $P(x \in t) \in \{0, 1\}$. $I$ is the set of all possible items.*

**Definition 2** *An* uncertain transaction $t$ *is a transaction that contains uncertain items. A transaction database $T$ containing uncertain transactions is called an* uncertain transaction database.

An uncertain transaction $t$ is represented in an uncertain transaction database by the items $x \in I$ associated with an existential probability value [1] $P(x \in t) \in (0, 1]$. An example of an uncertain transaction databases is depicted in Figure 1. To interpret an uncertain transaction database we apply the *possible world* model. An uncertain transaction database generates *possible worlds*, where each world is defined by a fixed set of (certain) transactions. A possible world is instantiated by generating each transaction $t_i \in T$ according to the occurrence probabilities $P(x \in t_i)$. Consequently, each probability $0 < P(x \in t_i) < 1$ derives two possible worlds *per transaction*: One possible world in which $x$ exists in $t_i$, and one possible world where $x$ does not exist in $t_i$. Thus, the number of possible worlds of a database increases exponentially in both the number of transactions and the number of uncertain items contained in it. Each possible world $w$ is associated with a probability that that world exists, $P(w)$.

---

[1] If an item $x$ has an existential probability of zero, it does not appear in the transaction.

We assume that uncertain transactions are mutually independent. This assumption is reasonable in real world applications. Additionally, independence between items is often assumed in the literature [7,8]. This can be justified by the assumption that the items are observed independently. In this case, the probability of a world $w$ is given by:

$$P(w) = \prod_{t \in I}(\prod_{x \in t} P(x \in t) * \prod_{x \notin t}(1 - P(x \in t)))$$

In cases where this assumption does not hold and conditional probabilities are available they may be used in our methods.

*Example 1.* In the database of Figure 1, the probability of the world existing in which $t_1$ contains only items $A$ and $C$ and $t_2$ contains only item $D$ is $P(A \in t_1)*(1-P(B \in t_1))*P(C \in t_1)*(1-P(A \in t_2)*P(D \in t_2) = 1.0 \cdot 0.8 \cdot 0.5 \cdot 0.9 \cdot 1.0 = 0.36$. For simplicity we omit the consideration of other customers in this example.

### 1.2 Problem Definition

An itemset is a *frequent itemset* if it occurs in at least *minSup* transactions, where *minSup* is a user specified parameter. In uncertain transaction databases however, the support of an itemset is uncertain; it is defined by a discrete probability distribution function (p.d.f). Therefore, each itemset has a *frequentness probability*[2] – the probability that it is frequent. In this paper, we focus on the two distinct problems of efficiently calculating this p.d.f. and efficiently extracting all *probabilistic frequent itemsets*;

**Definition 3** *A* Probabilistic Frequent Itemset *(PFI) is an itemset with a frequentness probability of at least $\tau$.*

The parameter $\tau$ is the user specified minimum confidence in the frequentness of an itemset.

We are now able to specify the *Probabilistic Frequent Itemset Mining (PFIM) problem* as follows; Given an uncertain transaction database $T$, a minimum support scalar *minSup* and a frequentness probability threshold $\tau$, find all probabilistic frequent itemsets.

### 1.3 Contributions

We make the following contributions:

- We introduce the probabilistic Frequent Pattern Tree, or ProFP-Tree, which is the first FP-Tree type approach for handling uncertain or probabilistic data. This tree efficiently stores a probabilistic database and enables efficient extraction of itemset occurrence probabilities and database projections.

---
[2] Frequentness is the rarely used word describing the property of being frequent.

- We propose ProFPGrowth, an algorithm based on the ProFPTree which mines all itemsets that are frequent with a probability of at least $\tau$ without using expensive candidate generation.
- We present an intuitive and efficient method based on generating functions for computing the probability that an itemset is frequent, as well as the entire probability distribution function of the support of an itemset, in $O(|T|)$ time[3]. Using our approach, our algorithm has the same time complexity as the approach based on the Poisson Binomial Recurrence (denoted as *dynamic programming technique*) in [6], but it is much more intuitive and thus offers various advantages, as we will show.

The remainder of this paper is organized as follows; Section 2 surveys related work. In Section 3 we present the ProFP-Tree, explain how it is constructed and briefly introduce the concept of conditional ProFPTrees. Section 4 describes how probability information is extracted from a (conditional) ProFP-Tree. Section 5 introduces our generating function approach for computing the frequentness probability and the support probability distribution in linear time. Section 6 describes how conditional ProFPT-rees are built. Finally, Section 7 describes the ProFP-Growth algorithm by drawing together the previous sections. We present our experiments in Section 8 and conclude in Section 9.

## 2  Related Work

There is a large body of research on Frequent Itemset Mining (FIM) but very little work addresses FIM in uncertain databases [7,8,13]. The approach proposed by Chui et. al [8] computes the expected support of itemsets by summing all itemset probabilities in their U-Apriori algorithm. Later, in [7], they additionally proposed a probabilistic filter in order to prune candidates early. In [13], the UF-growth algorithm is proposed. Like U-Apriori, UF-growth computes frequent itemsets by means of the expected support, but it uses the FP-tree [11] approach in order to avoid expensive candidate generation. In contrast to our probabilistic approach, itemsets are considered frequent if the expected support exceeds *minSup*. The main drawback of this estimator is that information about the uncertainty of the expected support is lost; [7,8,13] ignore the number of possible worlds in which an itemset is frequent. [21] proposes exact and sampling-based algorithms to find likely frequent items in streaming probabilistic data. However, they do not consider itemsets with more than one item. The current state-of-the-art (and only) approach for probabilistic frequent itemset mining (PFIM) in uncertain databases was proposed in [6]. Their approach uses an Apriori-like algorithm to mine all probabilistic frequent itemsets and the poisson binomial recurrence to compute the support probability distribution function (SPDF). We provide a faster solution by proposing the first probabilistic frequent pattern growth approach (ProFP-Growth), thus avoiding expensive candidate generation and allowing us to perform PFIM in large databases. Furthermore, we use a more intuitive generating function method to compute the SPDF.

---

[3] Assuming *minSup* is a constant.

Existing approaches in the field of uncertain data management and mining can be categorized into a number of research directions. Most related to our work are the two categories "*probabilistic databases*" [5,16,17,4] and "*probabilistic query processing*" [9,12,20,18].

The uncertainty model used in our approach is very close to the model used for probabilistic databases. A probabilistic database denotes a database composed of relations with uncertain tuples [9], where each tuple is associated with a probability denoting the likelihood that it exists in the relation. This model, called "*tuple uncertainty*", adopts the possible worlds semantics [4]. A probabilistic database represents a set of possible "certain" database instances (worlds), where a database instance corresponds to a subset of uncertain tuples. Each instance (world) is associated with the probability that the world is "true". The probabilities reflect the probability distribution of all possible database instances. In the general model description [17], the possible worlds are constrained by rules that are defined on the tuples in order to incorporate object (tuple) correlations. The ULDB model proposed in [5], which is used in *Trio*[2], supports uncertain tuples with alternative instances which are called x-tuples. Relations in ULDB are called x-relations containing a set of x-tuples. Each x-tuple corresponds to a set of tuple instances which are assumed to be mutually exclusive, i.e. no more than one instance of an x-tuple can appear in a possible world instance at the same time. Probabilistic top-k query approaches [18,20,16] are usually associated with uncertain databases using the tuple uncertainty model. The approach proposed in [20] was the first approach able to solve probabilistic queries efficiently under tuple independency by means of dynamic programming techniques. Recently, a novel approach was proposed in [15] to solve a wide class of queries in the same time complexity, but in a more elegant and also more powerful way using generating functions. In our paper, we adopt the generating function method for the efficient computation of frequent itemsets in a probabilistic way.

## 3 Probabilistic Frequent-Pattern Tree (ProFP-tree)

In this Section we introduce a novel prefix-tree structure that enables fast detection of probabilistic frequent itemsets without the costly candidate generation or multiple database scans that plague Apriori style algorithms. The proposed structure is based on the frequent-pattern tree (FP-tree [11]). In contrast to the FP-tree, the ProFP-tree has the ability to compress uncertain and probabilistic transactions. If a dataset contains no uncertainty it reduces to the (certain) FP-Tree.

**Definition 4 (ProFP-tree)** *A probabilistic frequent pattern tree is composed of the following three components:*

1. **Uncertain item prefix tree**: *A root labelled "null" pointing to a set of prefix trees each associated with uncertain item sequences. Each node n in a prefix tree is associated with an (uncertain) item $a_i$ and consists of five fields:*

- $n.item$ denotes the item label of the node. Let $path(n)$ be the set of items on the path from root to $n$.
- $n.count$ is the number of *certain* occurrences of $path(n)$ in the database.
- $n.uft$, denoting "uncertain-from-this", is the set of transaction ids (*tids*). A transaction $t$ is contained in uft *if and only if* $n.item$ is uncertain in $t$ (i.e. $0 < P(n.item \in t) < 1$) and $P(path(n) \subseteq t) > 0$.
- $n.ufp$, denoting "uncertain-from-prefix", is a set of transaction ids. A transaction $t$ is contained in ufp *if and only if* $n.item$ is certain in $t$ ($P(n.item \in t) = 1$) and $0 < P(path(n) \subseteq t) < 1$.
- $n.node-link$ links to the next node in the tree with the same item *label* if there exists one.

2. **Item header table**: This table maps all items to the first node in the Uncertain item prefix tree
3. **Uncertain-item lookup table**: This table maps $item, tid$ pairs to the probability that item appears in $t_{tid}$ for each transaction $t_{tid}$ contained in a uft of a node $n$ with $n.item = item$.

The two sets, *uft* and *ufp*, are specialized fields required in order to handle the existential uncertainty of itemsets in transactions associated with $path(n)$. We need two sets in order to distinguish where the uncertainty of an itemset (path) comes from. Generally speaking, the entries in $n.uft$ are used to keep track of existential uncertainties where the uncertainty is caused by $n.item$, while the entries in $ufp$ keep track of uncertainties of itemsets caused by items in $path(n) - n.item$ but where $n.item$ is certain.

Figure 2 illustrates the ProFP-tree of our example database of Figure 1. Each node of the *uncertain item prefix tree* is labelled by the field *item*. The labels next to the nodes refer to the node fields *count: uft ufp*. The dotted lines denote the *node-links*.

The ProFP-tree has the same advantages as a FP-tree, in particular: It avoids repeatedly scanning the database since the uncertain item information is efficiently stored in a compact structure. Secondly, multiple transactions sharing identical prefixes can be merged into one with the number of certain occurrences registered by *count* and the uncertain occurrences reflected in the transaction sets $uft$ and $ufp$.

### 3.1 ProFP-Tree Construction

For further illustration, we refer to our example database of Figure 1 and the corresponding ProFP-tree in Figure 2. We assume that the (uncertain) items in the transactions are lexicographically ordered, which is required for prefix tree construction.

We first create the root of the uncertain item prefix tree labelled "*null*". Then we read the uncertain transactions one at a time. While scanning the first transaction $t_1$, the first branch of the tree can be generated leading to the first path composing entries of the form (*item,count,uft,ufp,node-link*). In our example, the first branch of the tree is built by the following path:

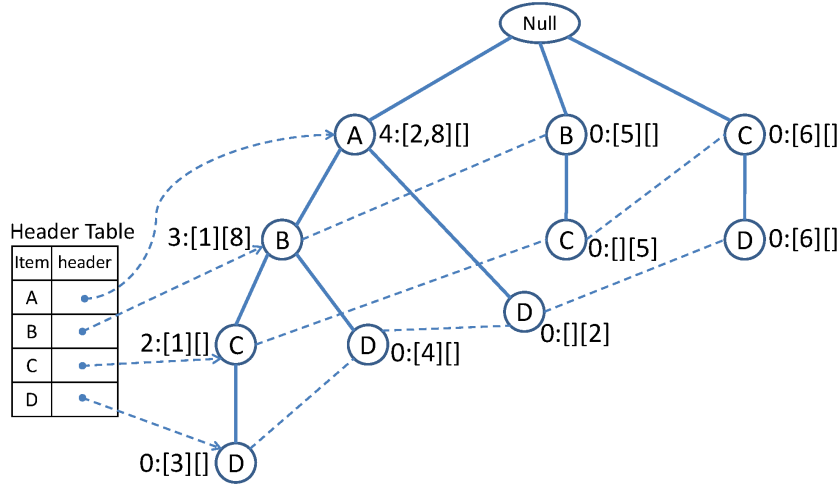

(a) *Uncertain item prefix tree* with *item header table*.

| $(1,B) \to 0.2$ | $(1,C) \to 0.5$ | $(2,A) \to 0.1$ |
| $(3,D) \to 0.4$ | $(4,D) \to 0.5$ | $(5,B) \to 0.1$ |
| $(6,C) \to 0.1$ | $(6,D) \to 0.5$ | $(8,A) \to 0.5$ |

(b) Uncertain-item lookup table.

**Fig. 2.** *ProFPTree* generated from the uncertain transaction database given in Figure 1.

$<root,(A,1,[],[],null),(B,0,[1],[],null),(C,0,[1],[],null)>$.

Note that the entry "1" in the field *uft* of the nodes associated with B and C indicate that item B and C are uncertain in $t_1$.

Next, we scan the second transaction $t_2$ and update the tree structure accordingly. The itemset of transaction $t_2$ shares its prefix with the previous one, therefore we follow the existing path in the tree starting at the root. Since the first item in $t_2$ is existentially uncertain, i.e. it exists in $t_2$ with a probability of 0.1, *count* of the first node in the path is not incremented. Instead, the current transaction $t_2$ is added to *uft* of this node. The next item in $t_2$ does not match with the next node on the path and, thus, we have to build a new branch leading to the leaf node $N$ with the entry $(D,0,[],[2],null)$. Although item D is existentially certain in $t_2$ *count* of $N$ is initialized with zero, because the itemset A,D associated with the path from the root to node $N$ is existentially uncertain in $t_2$ due to the existential uncertainty of item A. Hence, we add transaction $t_2$ to the *uncertain-from-prefix (ufp)* field of $n$. The resulting tree is illustrated in Figure 3(a).

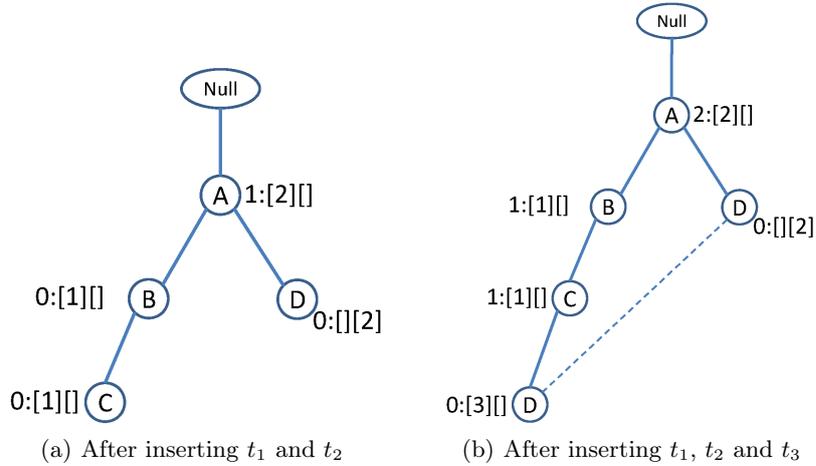

(a) After inserting $t_1$ and $t_2$  (b) After inserting $t_1$, $t_2$ and $t_3$

**Fig. 3.** *Uncertain item prefix tree* after insertion of the first transactions.

The next transaction to be scanned is transaction $t_3$. Again, due to matching prefixes we follow the already existing path $<A,B,C>$[4] while scanning the (uncertain) items in $t_3$. The resulting tree is illustrated in Figure 3(b). Since the first item A is existentially certain, *count* of the first node in the prefix path is incremented by one. The next items, item B and C, are registered in the tree in the same way by incrementing the *count* fields. The rational for these *count* increments is that the corresponding itemsets are existentially certain in $t_3$. The final item D is processed by adding a new branch below the node C leading to a new leaf node with the fields: $(D,0,[3],[],ptr)$, where the link *ptr* points to the next node in the tree labelled with item label D. Since item D is existentially uncertain in $t_3$ the *count* field is initialized with 0 and $t_3$ is registered in the $uft$ set. The *uncertain item prefix tree* is completed by scanning all remaining transactions in a similar fashion.

The ProFP-tree construction algorithm is shown in Algorithm 1.

### 3.2 Construction Analysis

The construction of the ProFP-tree requires a single scan of the uncertain transaction database $\mathcal{T}$. For each processed transaction we must follow and update or construct a single path of the tree, of length equal to the number of items in the corresponding transaction. Therefore the ProFP-tree is constructed in linear time w.r.t. to size of the database.

Since the ProFP-tree is based on the original FP-tree, it inherits its compactness properties. In particular, the size of a ProFP-tree is bounded by the

---
[4] To simplify matters, we use the *item* fields to address the nodes in a path, just for illustration.

---

**Algorithm 1** ProFP-Tree ConstructionCreation.

---

**input:** An uncertain transaction Database $\mathcal{T}$ with lexicographically ordered items, and a minimum support threshold *minSup*.
**Output:** A *probabilistic frequent pattern tree* (ProFP-Tree).

**Method:**
  Create the (null) root of an *uncertain item prefix tree* $T$;
  Initialize an empty *item header table* (*iht*);
  Initialize an empty *uncertain-item lookup table* (*ult*);
  for each uncertain transaction $t_i \in \mathcal{T}$
    Build a string $<it_1, \cdots, it_n>$ of tuples $it_j=(item,prob)$,
    where the field *item* identifies a(n) (un)certain item of $t_i$
    and the field *prob* denotes the probability $P(it_j.item \in t_i)$.
    Call *insert-transaction*($<it_1, \cdots, it_n>,i,T.root,0$)

***insert-transaction***(*transaction,i,node,u_flag*)
while *it*:= *transaction.get_next_item*() not *null* do
  if *node* has a child $N$ with $N.item = it.item$, then
    call *update-node-entries*(*it,i,N,u_flag*); //follow exist. path
  else //create new branch:
    create new child $N$ of $T$;
    call *update-node-entries*(*it,i,N,u_flag*);
    if *it.item* not in *iht* then
      insert (*it.item,ptr(N)*) into *iht*;
    else
      insert node $N$ into the link list associated with *it.item*;
  //update *uncertain-item lookup table*
  if *it.prob*<1.0 then
    insert (*i,it.item,it.prob*) into *ult*;
  *node*:= N;

***update-node-entries***(*it,i,N,u_flag*)
  if *it.prob*=1.0, then
    if *u_flag*=0 then
      increment $N.count$ by 1;
    else //*u_flag*=1
      insert $i$ into $N.ufp$;
  else
    insert $i$ into $N.uft$;
    set *u_flag*:=1;

---

overall occurrences of the (un)certain items in the database and its height is bounded by the maximal number of (un)certain items in a transaction. For any transaction $t_i$ in $\mathcal{T}$, there exists exactly one path in the *uncertain item prefix tree* starting below the *root* node. Each item in the transaction database can create no more than one node in the tree and the height of the tree is bounded by the number of items in a transaction (path). Note that as with the FP-Tree, the compression is obtained by sharing common prefixes.

We now show that the values stored at the nodes do not affect the bound on the size of the tree. In particular, in the following Lemma we bound the *uncertain-from-this (uft)* and *uncertain-from-prefix (ufp)* sets.

**Lemma 5** *Let $T$ be the* uncertain item prefix tree *generated from an uncertain transaction database $\mathcal{T}$. The total space required by all the transaction-id sets (*uft *and* ufp*) in all nodes in $T$ is bounded by the the total number of uncertain occurrences[5] in $\mathcal{T}$.*

The rational for the above lemma is that each occurrence of an uncertain item (with existence probability in $(0, 1)$) in the database yields at most one transaction-id entry in one of the transaction-id sets assigned to a node in the tree. In general there are three update possibilities for a node $N$: If the current item and all prefix items in the current transaction $t_i$ are certain, there is no new entry in $uft$ or $ufp$ as *count* is incremented. $t_i$ is registered in $N.uft$ if and only if $N.item$ is existentially uncertain in $t_i$ while $t_i$ is registered in $N.ufp$ if and only if $N.item$ is existentially certain in in $t_i$ but at least one of the prefix items in $t_i$ is existentially uncertain. Therefore each occurrence of an item in $\mathcal{T}$ leads to either a count increment or a new entry in $uft$ or $ufp$.

Finally, it should be clear that the size of the uncertain item lookup table is bounded by the number of uncertain (non zero and non 1) entries in the database.

In this section we showed that the ProFP-Tree inherits the compactness of the original FP-Tree. In the following Section we show that the information stored in the ProFP-tree suffices to retrieve all probabilistic information required for PFIM, thus proving completeness.

## 4 Extracting Certain and Uncertain Support Probabilities

Unlike the (certain) FP-Growth approach where extracting the support of an itemset $X$ is easily achieved by summing the support counts along the node-links for $X$ in a suitable conditional ProFPTree, we are interested in the support distribution of $X$ in the probabilistic case. Before we can compute this however, we first require both the number of certain occurrences as well as the probabilities $0 < P(X \in t_i) < 1$. Both can be efficiently obtained using the ProFP-Tree as follows:

To obtain the certain support of an item $x$, follow the node-links from the header table and accumulate both the counts and the number of transactions in which $x$ is uncertain-from-prefix. The latter is counted since we are interested in the support of $x$ and by construction, transactions in $ufp$ are known to be certain for $x$. To find the set of transaction ids in which $x$ is uncertain, follow the node-links and accumulate all transactions that are in the uncertain-from-this ($uft$) list.

---
[5] Entries in transactions with an existential probability in $(0, 1)$.

*Example 2.* By traversing the node-list, we can calculate the certain support for item $C$ in the $ProFP$-Tree in Figure 2 as follows: $2 + |\emptyset| + |\{t_5\}| + |\emptyset| = 3$. Note there is one transaction in which $C$ is uncertain-from-prefix ($t_5$). Similarly, we find that the only transactions in which $C$ is uncertain are $t_1$ and $t_6$. The exact appearance probabilities in these transactions can be obtained from the uncertain-item lookup table. By comparing this to Figure 1 we see that the tree allows us to obtain the correct certain support and the transaction ids where $C$ is uncertain.

To compute the support of an itemset $X = \{a, ..., k\}$, we use the conditional tree for items $b, ..., k$ and extract the certain support and uncertain transaction ids for $a$. Since it is somewhat involved, we defer the construction of conditional ProFP-Trees to Section 6. By using the conditional tree, the above method provides the certain support of $X$ and the exact set of transaction ids in which $X$ is uncertain ($utids$). To compute the probabilities $P(X \in t_i) : t_i \in utids$ we use the independence assumption and multiply, for each $x \in X$ the probability that $x$ appears in $t_i$. Recall that the probability that $X$ appears in $t_i$ is an $O(1)$ lookup in the uncertain-item lookup table. Recall that if additional information is given on the dependencies between items, this can be incorporated here.

We have now described how the certain support and all probabilities $P(X \in t) : X\ uncertain\ in\ t$ can be efficiently computed from the ProFPTree (Algorithm 2). Section 5 shows how we use this information to calculate the support distribution of $X$.

## 5 Efficient Computation of Probabilistic Frequent Itemsets

This section presents our linear-time technique for computing the probabilistic support of an itemset using generating functions. The problem is as follows:

**Definition 6** *Given a set of $N$ mutually independent but not necessarily identical Bernoulli (0/1) random variables $P(X \in t_i)$, $1 \leq i \leq N$, compute the probability distribution of the random variable $Sup = \sum_{N}^{i=1} X_i$*

A naive solution would be to count for each $0 \leq k \leq N$ all possible worlds in which exactly $k$ items contain $X$ and accumulate the respective probabilities. This approach however, shows a complexity of $O(2^N)$. In [6] an approach has been proposed that achieves an $O(N)$ complexity using Poisson Binomial Recurrence. Note that $O(N)$ time is asymptotically optimal in general, since the computation involves at least $O(N)$ computations, namely $P(X \in t_i) \forall 1 \leq i \leq N$. In the following, we propose a different approach that, albeit having the same linear asymptotical complexity, has other advantages.

**Algorithm 2** Extract Probabilities for an itemset.

*//calcuate the certain support and the uncertain transaction ids of an item*
*//derived from a PFP-Tree*
extract(*item*,*ProFP − Tree tree*)
  $certSup = 0$; $uncertainSupTids = \emptyset$;
  for each *ProFPNode* in *tree* reachable
  from header table[*item*]
    $certSupp+ = n.certSupp$;
    $certSupp+ = |n.ufp|$;
    $uncertainSupTids = uncertainSupTids \cup n.uft$;
  return *certSupp*,*uncertainSupTids*;

*//calculate the existential probabilities of an itemset*
calculateProbabilities(*itemset, uncertainSupTids*)
  $probabilityVector = \emptyset$;
  for ($t \in uncertainSupTids$)
    $p = \Pi_{i\ in\ itemset} uncertainItemLookupTable[i,t]$;
    $probabilityVector$.add($p$);
  return *probabilityVector*;

### 5.1 Efficient Computation of Probabilistic Support

We apply the concept of generating functions as proposed in the context of probabilistic ranking in [15]. Consider the function: $\mathcal{F}(x) = \prod_{i=1}^{n}(a_i + b_i x)$. The coefficient of $x^k$ in $\mathcal{F}(x)$ is given by: $\sum_{|\beta|=k} \prod_{i:\beta_i=0} a_i \prod_{i:\beta_i=1} b_i$, where $\beta = \langle \beta_1, ..., \beta_N \rangle$ is a Boolean vector, and $|\beta|$ denotes the number of 1's in $\beta$.

Now consider the following generating function:

$$\mathcal{F}^i = \prod_{t \in \{t_1,...t_i\}} (1 - P(X \in t) + P(X \in t) \cdot x) = \sum_{j \in \{0,...,i\}} c_j x^j$$

The coefficient $c_j$ of $x^j$ in the expansion of $\mathcal{F}^i$ is exactly the probability that $X$ occurs in exactly $j$ if the first $i$ transactions; that is, the probability that the support of $X$ is $j$ in the first $i$ transactions. Since $\mathcal{F}^i$ contains at most $i + 1$ nonzero terms and by observing that

$$\mathcal{F}^i = \mathcal{F}^{i-1} \cdot (1 - P(X \in t_i) + P(X \in t_i)x)$$

we note that $\mathcal{F}^i$ can be computed in $O(i)$ time given $\mathcal{F}^{i-1}$. Since $\mathcal{F}^0 = 1x^0 = 1$, we conclude that $\mathcal{F}^N$ can be computed in $O(N^2)$ time. To reduce the complexity to $O(N)$ we exploit that we only need to consider the coefficients $c_j$ in the generating function $\mathcal{F}^i$ where $j < minSup$, since:

- The frequentness probability of $X$ is defined as $P(X\ is\ frequent) = P(Sup(X) \geq minSup)) = 1 - P(Sup(X) < minSup) = 1 - \sum_{j=0}^{minSup-1} c_j$

– A coefficient $c_j$ in $\mathcal{F}^i$ is independent of any $c_k$ in $\mathcal{F}^{i-1}$ where $k > j$. That means in particular that the coefficients $c_k$, $k \geq minSup$ are not required to compute the $c_i$, $i < minSup$.

Thus, keeping only the coefficients $c_j$ where $j < minSup$, $\mathcal{F}^i$ contains at most $minSup$ coefficients, leading to a total complexity of $O(minSup \cdot N)$ to compute the frequentness probability of an itemset.

*Example 3.* As an example, consider itemset $\{A, D\}$ in the running example database in Figure 1. Using the *ProFP-Tree* (c.f. Figure 2(a)), we can efficiently extract, for each transaction $t_i$, the probability $P(\{A, D\} \in t_i)$, where $0 < P(\{A, D\} \in t_i) < 1$ and also the number of certain occurrences of $\{A, D\}$. Itemset $\{A, D\}$ certainly occurs in no transaction and occurs in $t_2, t_3$ and $t_4$ with a probability of 0.1, 0.4 and 0.5 respectively. Let minSup be 2:

$$\mathcal{F}^1 = \mathcal{F}^0 \cdot (0.9 + 0.1x) = 0.1x^1 + 0.9x^0$$
$$\mathcal{F}^2 = \mathcal{F}^1 \cdot (0.6 + 0.4x) = 0.04x^2 + 0.42x^1 + 0.54x^0 \stackrel{*}{=} 0.42x^1 + 0.54x^0$$
$$\mathcal{F}^3 = \mathcal{F}^2 \cdot (0.5 + 0.5x) = 0.21x^2 + 0.48x^1 + 0.27x^0$$
$$\stackrel{*}{=} 0.48x^1 + 0.27x^0$$

Thus, $P(sup(\{A, D\}) = 0) = 0.27$ and $P(sup(\{A, D\}) = 1) = 0.48$. We get that $P(sup(\{A, D\}) \geq 2) = 0.25$. Thus, {A,D} is not returned as a frequent itemset if $\tau$ is greater than 0.25. Equations marked by a * exploit that we only need to compute the $c_j$ where $j < minSup$.

Note that at each iteration of computing $\mathcal{F}^i$, we can check whether $1 - \sum_{i<minSup} c_i \geq \tau$ and if that is the case, we can stop the computation and conclude that the respective itemset (for which $\mathcal{F}$ is the generating function) is frequent. Intuitively, the reason is that if an itemset $X$ is already frequent considering the first $i$ transactions only, $X$ will still be frequent if more transactions are considered. This intuitive pruning criterion corresponds to the pruning criterion proposed in [6] for the Poisson Binomial Recurrence approach.

We remark that the generating function technique can be seen as a variant of the Poisson Binomial Recurrence. However, using generating functions instead of the complicated recursion formula gives us a much cleaner view on the problem. In addition, using generating functions, the support probability density function (sPDF) can be updated easily if a transaction $t_i$ changes its probability of containing an itemset $X$. That is, if the probability $p = P(X \in t_i)$ changes to $p'$, then we can simply obtain the expanded polynomial from the old sPDF and divide it by $px + (1 - p)$ (using polynomial division) to remove the effect of $t_i$ and multiply $p'x + (1-p')$ to incorporate the new probability of $t_i$ containing $X$. That is, $\mathcal{F}^{i'}(x) = \mathcal{F}^i(x) : (px + 1 - p) \times (p'x + 1 - p')$, where $\mathcal{F}^{i'}$ is the generating function of the sPDF of $X$ in the changed database containing $t'_i$.

## 6 Extracting Conditional ProFP-Trees

This section describes how conditional ProFP-Trees are constructed from other (potentially conditional) ProFP-Trees. The method for doing this is more involved than the analogous operation for the certain FPGrowth algorithm, since we must ensure that the information capturing the source of the uncertainty remains correct. That is, whether the uncertainty at that node comes from the prefix or from the present node. Recall from Section 4 that this is required in order to extract the correct probabilities from the tree. A conditional ProFP-Tree for itemset $X$ ($tree_X$) is equivalent to a ProFP-Tree built on only those transactions in which $X$ occurs with a non-zero probability. In order to generate a conditional ProFP-Tree for itemset $X \cup i$ ($tree_{X \cup i}$) where $i$ occurs lexicographically prior to any item in $X$, we first begin with the conditional ProFP-Tree for $X$. When $X = \emptyset$, $tree_X$ is simply the complete ProFP-Tree. We construct $tree_{X \cup i}$ by propagating the values at the nodes with $item = i$ upwards and accumulating these at the nodes closer to the root as listed in Algorithm 3. Let $N_i$ be the set of nodes with $item = i$ (These are obtained by following the links from the header table). The values for every node $n$ in the resulting conditional tree $tree_{X \cup i}$ are calculated as follows:

- $n.count = \sum_{n_i \in N_i} n_i.count$ since these represent certain transactions.
- $n.uft = \cup n_i.uft | n_i \in N_i$ since we are conditioning on an item that is uncertain in these transactions and hence any node in the final conditional tree will also be uncertain for these transactions.
- When collecting transactions for $n$ that are uncertain from the prefix (i.e. $t \in ufp$), we must determine whether the item $n.item$ caused this uncertainty. If the corresponding node in $tree_X$ contained transaction $t$ in $ufp$, then $t$ is also in $n.ufp$ ($n.item$ was not uncertain in $t$). If $n.item$ was uncertain in $t$, then the corresponding node in $tree_X$ would have $t$ listed in $uft$ and this must also remain the case for the conditional tree. If $t \in n.ufp$ is neither in the corresponding $ufp$ nor $uft$ in $tree_X$, then it must be certain for $n.item$ and $n.count$ is incremented. Using this approach, we can avoid storing the set of transactions for which an item is certain. This is a key idea in our ProFP-Tree.

## 7 ProFP-Growth Algorithm

We have now described the three fundamental operations of the ProFP-Growth Algorithm; building the ProFPTree (Section 3); efficiently extracting the certain support and uncertain transaction probabilities from it (Section 4); calculating the frequentness probability and determining whether an item(set) is a probabilistic frequent itemset (Section 5); and construction of the conditional ProFPTrees (Section 6). Together with the fact that probabilistic frequent itemsets possess an antimonotonicity property (Lemma 17 in [6]), we can use a similar approach to the certain FPGrowth algorithm to mine all probabilistic frequent itemsets. Since, in principle, this is not substantially different from substituting the corresponding steps in FP-Growth, we omit further details.

**Algorithm 3** Construction of a conditional ProFP-Tree $tree_{X \cup i}$ by 'extracting' item $i$ from the conditional ProFP-Tree for itemset $X$.

*//Accumulates transactions for nodes when propagating up the values*
*//from a node being extracted.*
class *Accumulator*
  $count = 0$; $uft = \emptyset$; $ufp = \emptyset$;
  $orig\_ufp = $ the original ufp list
  add($ProFPNode\ n$)
    $count+ = n.count$;
    $uft = uft \cup n.uft$;
    for ($t \in n.ufp$)
      if ($orig\_ufp.contains(t)$) $ufp = ufp \cup t$;
      else if ($orig\_uft.contains(t)$) $uft = uft \cup t$;
      else $count ++$;

buildConditionalProFPTree($ProFPTree\ tree_X$, $item\ i$) returns $tree_{X \cup i}$
  $tree_{X \cup i} = $clone of the subtree of $tree_X$ reachable from header table for $i$;
  associate an *Accumulator* with each node in $tree_{X \cup i}$ and set $orig\_ufp$;
  propagate($tree_{X \cup i}, i$);
  set the $certSup, uft, ufp$ values of nodes in $tree_{X \cup i}$ to those in the
    corresponding *Accumulator*s;

propagate($ProFPTree\ tree$, $item\ i$)
  for($ProFPNode\ n$ accessible from header table for $i$)
    $ProFPNode\ cn = n$;
    while($cn.parent \neq null$)
      call add($n$) on Accumulator for $cn$;
      $cn = cn.parent$;

## 8 Experimental Evaluation

In this section, we present performance experiments using our proposed *ProFP-Growth* algorithm and compare the results to the Apriori-based solution (denoted as *ProApriori*) presented in [6]. We also analyze how various database characteristics and parameter settings affect the performance of $ProFP-Growth$.

All experiments were performed on an Intel Xeon with 32 GB of RAM and a 3.0 GHz processor. For the first set of experiments, we used artificial datasets with a variable number of transactions and items. Each item $x$ has a probability $P_1(x)$ of appearing for certain in a transaction, and a probability $P_0(x)$ of not appearing at all in a transaction. With a probability $1 - P_0(x) - P_1(x)$ item $x$ is therefore uncertain in a transaction. In this case, the probability that $x$ exists in a transaction is picked randomly from a uniform $(0, 1)$ distribution.

For our scalability experiments, we scaled the number of items and transactions and chose $P_0(x) = 0.5$ and $P_1(x) = 0.2$ for each item. We measured the run time required to mine all probabilistic frequent itemsets that have a minimum support of 10% of the database size with a probability of a least $\tau = 0.9$.

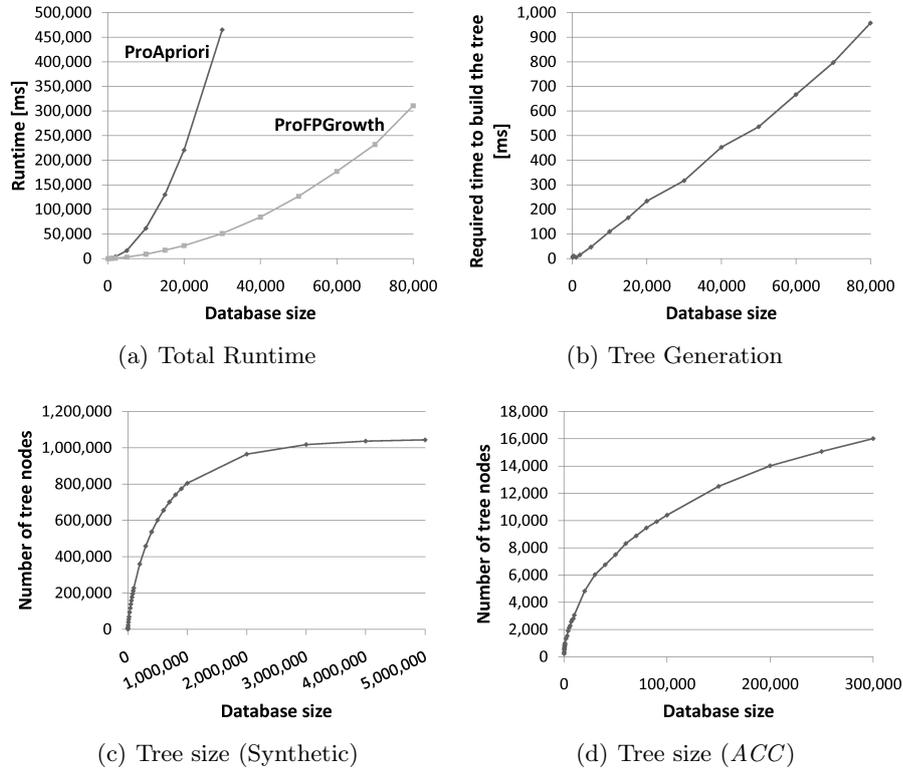

(a) Total Runtime  (b) Tree Generation

(c) Tree size (Synthetic)  (d) Tree size ($ACC$)

**Fig. 4.** Scalability w.r.t. the number of transactions.

### 8.1 Number of Transactions

We scaled the number of transactions and used 20 items. The results can be seen in Figure 4(a). In this setting, our approach significantly outperforms *ProApriori* [6]. The time required to build the *ProFP-Tree* w.r.t. the number of transactions is depicted in Figure 4(b). The observed linear run time indicates a constant time required to insert transactions into the tree. This is expected since the maximum height of the *ProFP-Tree* is equal to the number of items. Finally, we evaluated the size of the *ProFP-Tree* for this experiment, shown in Figure 4(c). The number of nodes in the *ProFP-Tree* increases and then plateus as the number of transactions increases. This is because new nodes have to be created for those transaction where a suffix of the transaction is not yet contained in the tree. As the number of transactions increases, the overlap between transaction prefixes increases, requiring fewer new nodes to be created. It is expected that this overlap increases faster if the items are correlated. Therefore, we evaluate the size of the *ProFP-Tree* on subsets of the real-world dataset *accidents*[6], denoted by $ACC$. It consists of $340,184$ transactions and a reduced number of 20 items whose occurrences in transactions were randomized; with a probability of 0.5, each item appearing for certain in a transaction was assigned a value drawn from a uniform distribution in $(0, 1]$. We varied the number of transactions from $ACC$ up to the first $300,000$. As can be seen in Figure 4(d), there is more overlap between transactions since the growth in the number of nodes used is slower (compared to Figure 4(c)).

---

[6] The *accidents* dataset [10] was derived from the Frequent Itemset Mining Dataset Repository (http://fimi.cs.helsinki.fi/data/)

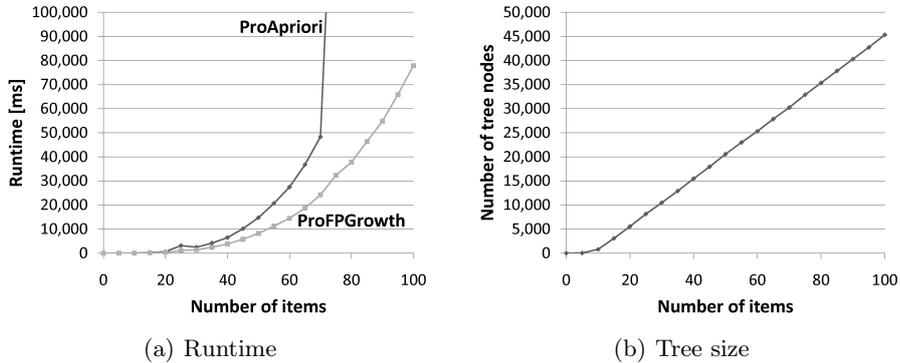

(a) Runtime  (b) Tree size

**Fig. 5.** Scalability with respect to the number of items.

### 8.2 Number of Items

Next, we scaled the number of items using $1,000$ transactions. The run times for 5 to 100 items can be seen in Figure 5(a), which shows the expected exponential runtime inherent in FIM problems. It can be clearly seen that the *ProFP-Growth* approach vastly outperforms *ProApriori*.

Figure 5(b) shows the number of nodes used in the *ProFP-Tree*. Except for very few items, the number of nodes in the tree grows linearly.

### 8.3 Effect of Uncertainty and Certainty

In this experiment, we set the number of transactions to $1,000$ and the number of items to 20 and varied the parameters $P_0(x)$ and $P_1(x)$.

For the experiment shown in Figure 6(a), we fixed the probability that items are uncertain $(1 - P_0(x) - P_1(x))$ at 0.3 and successively increased $P_1(x)$ from 0 (which means that no items exist for certain) to 0.7. It can be observed that the number of nodes initially increases. This is what we would expect, since more items existing in the database increases the nodes required. However, as the number of certain items increases, an opposite effect reduces the number of nodes in the tree. This effect is caused by the increasing overlap of the transactions – in particular, the increased number and length of shared prefixes. When $P_1(x)$ reaches 0.7 (and thus $P_0(x) = 0$), each item is contained in each transaction with a probability greater than zero, and thus all transactions contain the same

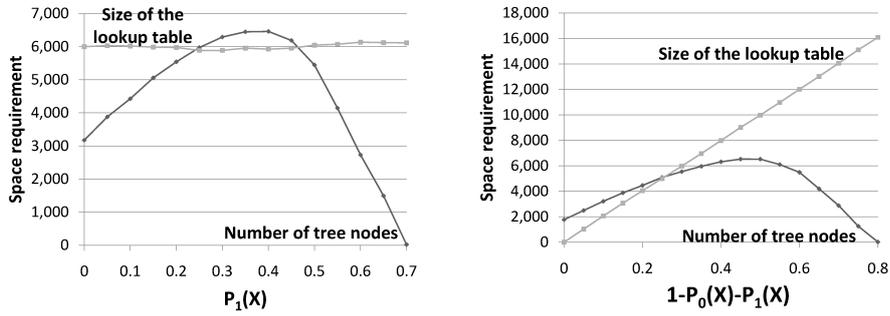

(a) Varying the probability of certain occurrences while keeping uncertain occurrences fixed.

(b) Varying the probability of uncertain occurrences while keeping certain occurrences fixed.

**Fig. 6.** Effect of certainty and uncertainty on the ProFP-Tree size and uncertain item lookup table.

items with non-zero probability. In this case, the *ProFP-Tree* degenerates to a linear list containing exactly one node for each item. Note that the size of the uncertain item lookup table is constant, since the expected number of uncertain items is constant at $0.3 \cdot |\mathcal{T}| \cdot |I| = 0.3 \cdot 1{,}000 \cdot 20 = 6{,}000$.

In Figure 6(b) we fixed $P_1(x)$ at 0.2 and successively decreased $P_0(x)$ from 0.8 to 0, thus increasing the probability that items are uncertain from 0 to 0.8. We see a similar pattern as in Figure 6(a) for the number of nodes, for similar reasons. As expected here, the size of the lookup table increases as the number of uncertain items increases.

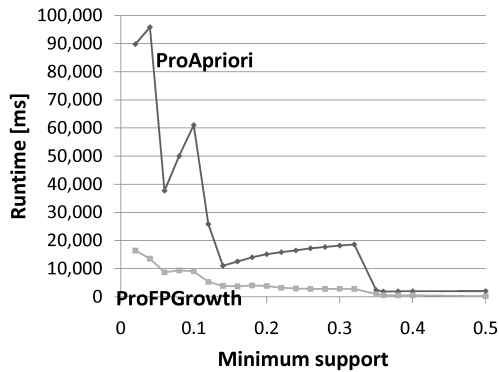

**Fig. 7.** Effect of *minSup*

### 8.4 Effect of *minSup*

Here, we varied the minimum support threshold *minSup* using an artificial database of 10,000 transactions and 20 items. Figure 7 shows the results. For low values of *minSup*, both algorithms have a high run time due to the large number of probabilistic frequent itemsets. It can be observed that *ProFP-Growth* significantly outperforms *ProApriori* for all settings of *minSup*.

## 9 Conclusion

The Probabilistic Frequent Itemset Mining (PFIM) problem is to find itemsets in an uncertain transaction database that are (highly) likely to be frequent. This problem has two components; efficiently computing the support probability distribution and frequentness probability, and efficiently mining all probabilistic frequent itemsets. To solve the first problem in linear time, we proposed a novel method based on generating functions. To solve the second problem, we proposed the first probabilistic frequent pattern tree and pattern growth algorithm. We demonstrated that this significantly outperforms the current state of the art approach to PFIM.